\documentclass[12pt]{article}
\usepackage{graphics}
\usepackage{dcolumn}
\usepackage{amssymb}
\newcommand{\bea}{\begin{eqnarray}}
\newcommand{\eea}{\end{eqnarray}}
\newcommand{\be}{\begin{equation}}
\newcommand{\ee}{\end{equation}}

\def\be{\begin{eqnarray}}
\def\ee{\end{eqnarray}}
\def\bd{\begin{displaymath}}
\def\ed{\end{displaymath}}

\def\NP{Nucl. Phys. }
\def\PR{Phys. Rev. }
\def\PRL{Phys. Rev. Lett. }
\def\PL{Phys. Lett. }

\begin{document}
\title{Spectroscopic factors for alpha decay in the $N_pN_n$ scheme}
\author{Madhubrata Bhattacharya$^a$, Subinit Roy$^b$ 
and G. Gangopadhyay$^a$
\footnote{Corresponding Author}
\footnote{email: ggphy@caluniv.ac.in}\\
$^a$Department of Physics, University of Calcutta\\
92, Acharya Prafulla Chandra Road, Kolkata-700 009, India\\
$^b$Saha Institute of Nuclear Physics, Block AF, Sector 1\\
Kolkata- 700 064, India}
\date{}
\maketitle
\begin{abstract}
Lifetime values for alpha decay in even-even nuclei with $Z=84-98$ and 
$N=128-152$
have been calculated in the superasymmetric fission model. The interaction 
between the alpha particle and the daughter nucleus has been formed in 
the double folding approach using a density dependent NN interaction. The
densities have been obtained  using the Relativistic Mean Field formalism.  
The spectroscopic factors for the decays have been deduced and are shown to
vary smoothly as a function of effective numbers of valence nucleons,
$N_p$ and $N_n$ chosen with a suitable core. The implication of such a smooth behaviour has been
discussed.

\end{abstract}
                                                                                


\vskip 0.5cm

Simplified parametrization of various nuclear quantities may be obtained if 
the quantities are plotted as a function of $N_pN_n$, the product of effective 
number of valance particles (or holes)\cite{Casten1}. Various quantities such
as deformation and B(E2) values\cite{Casten3,Foy,Zhao}, rotational 
moments of inertia in low spin states in the rare earth region\cite{Saha}, 
ground band energy systematics \cite{Saha1}, core cluster decomposition in the 
rare earth region\cite{Buck} and properties of excited states 
\cite{Casten2,yoon}
have been found to follow certain simple trends when expressed as a function
of simple product of $N_p$ and $N_n$ or certain simple functions of the above 
two numbers. Essentially this simple functions are seen to represent
$n-p$ interaction and bear smooth relationships with the observables. 

It has often been pointed out that in view of the change in magic number and 
shell structure in various mass regions, the conventional counting of valence 
protons and neutrons may be inadequate. Extraction of effective number of 
valence particles in the  $N_pN_n$ scheme may significantly improve
the predictive capability of the scheme as well as point to the emergence of
new shell structure in different mass regions\cite{Casten1,Casten4}. 
For example, Zhao {\em et al.}\cite{Casten4} have used the concept of effective valence number of protons 
and neutrons to study the breakdown of the shell gap at Z=64 in even-even, odd 
and odd-odd nuclei. Wolf and Casten\cite{Wolf} also have studied the shell gap 
at $Z=64$ for $N_pN_n$ scheme.

Alpha decay in the heavy actinide nuclei presents a unique opportunity of 
testing various nuclear structure theories. It is known to take 
place through tunnelling of the potential barrier by the alpha particle.
The spectroscopic factor in $\alpha$-decay  was introduced to incorporate
the preformation probability. It contains the nuclear structure effects, 
and may be thought as the overlap between the actual ground state configuration 
of the parent and the configuration described by one $\alpha$-particle coupled 
to the ground state of the daughter. Thus the tunnelling probability of the 
alpha particle is a 
quantity which is expected to be crucially dependent on the $n-p$ interaction
and the quantity spectroscopic factor may show a certain correlation when expressed as a function of $N_p$ and $N_n$.

In the present work, we have followed the microscopic superasymmetric fission 
model. The interaction potential between the alpha particle and the daughter 
nucleus has been obtained in the double folding model by folding the densities
 of 
the alpha particle and the daughter nucleus using some suitable interaction. 
Usually the densities are obtained from phenomenological description. However,
microscopic densities obtained from mean field approaches may be expected to 
provide a better description of the densities and hence that of the process
of alpha decay.

Relativistic Mean Field (RMF) approach is now a standard tool in low energy
nuclear structure.
It has been able to explain different features
of stable and exotic nuclei like ground state binding energy, deformation,
radius, excited states, spin-orbit splitting, neutron halo, etc.
For a detailed discussion, readers are referred to the refs.
\cite{Serot,RMF1,Serot1}. Later related developments in studies of ground state
structure include incorporating the effect of the 
resonant continuum in drip line nuclei in both mean field \cite{rmfc} and 
Hartree Bogoliubov approach\cite{RHB}, study of new Lagrangian densities with
coupling between mesons\cite{prl}, etc.
It is well known that in nuclei far away from the
stability valley, the single particle level structure undergoes certain
changes in which the spin-orbit splitting plays an important role.
Being based on the Dirac Lagrangian density, RMF is particularly suited to
investigate these nuclei because it naturally incorporates the
spin degrees of freedom.

There exist different variations of the Lagrangian density as
well as a number of different parameterizations in RMF. Recently, a new Lagrangian
density has been proposed\cite{prl} which involves
self-coupling of the vector-isoscalar meson as well as coupling between the
vector-isoscalar meson and the vector-isovector meson. The corresponding
parameter set is called FSU Gold\cite{prl}. This Lagrangian density was earlier 
employed to obtain the proton nucleus interaction to successfully 
calculate the half life for proton radioactivity\cite{plb}. 
We have also applied it to study alpha radioactivity in superheavy 
nuclei with $A>282$ \cite{282} and cluster radioactivity\cite{cluster} in the 
same approach followed in the present work. This density seems very appropriate for a large mass region {\em viz}. medium mass to superheavy nuclei. In this 
work also, we have employed FSU Gold.

In the conventional RMF+BCS approach for even-even nuclei, the Euler-Lagrange
equations are solved under the assumptions of classical meson
fields, time reversal symmetry, no-sea contribution, etc. Pairing is introduced
under the BCS approximation. Since accuracy of the nuclear density is very
important in our calculation, we have solved the equations in co-ordinate space.
The strength of the zero range pairing force is taken as 300 MeV-fm for both
protons and neutrons. 

The microscopic density dependent M3Y interaction (DDM3Y) is obtained from a 
finite range nucleon interaction by introducing a density dependent factor. 
This class of interactions has been employed widely in the study of 
nucleon-nucleus as well as nucleus-nucleus scattering, calculation of proton 
radioactivity, etc.
In this work,
we have employed the exponential density dependent interaction DDM3Y1
\bea
v(r,\rho_1,\rho_2,E)=C(1+\alpha\exp{(-\beta(\rho_1+\rho_2)})
(1-0.002E)u^{M3Y}(r)\eea
used in Ref. \cite{Khoa} to study alpha-nucleus scattering. It uses the direct 
M3Y potential $u^{M3Y}(r)$ based on the $G$-matrix elements of the 
Reid\cite{Reid} NN potential. The weak energy dependence was introduced\cite{Khoa1} to 
reproduce the empirical energy dependence of the optical potential. The 
parameters used are the standard values {\em viz.} $C=0.2845$, $\alpha=3.6391$ 
and $\beta=2.9605$fm$^2$. Here $\rho_1$ and $\rho_2$ are the densities of 
the $\alpha$-particle and the daughter nucleus and $E$ is the energy per nucleon
of the $\alpha$-particle in MeV.  This interaction has been double folded
with the theoretical densities of alpha particle and the daughter nucleus
in their ground states using the code DFPOT\cite{dfpot}.  The assault frequency
has been calculated from the decay energy following Gambhir
{\em et al.}\cite{gambhir}.

We have studied the alpha decays between the ground states of even-even nuclei
for parents with $Z=84-98$ and $N=128-152$.
The spectroscopic factor has been calculated as the ratio of the calculated half
life to the experimentally observed value. Obviously it is expected to be less 
than unity. 
The results for the spectroscopic factors are presented in Table \ref{tab1}.
Theoretical calculations exist for spectroscopic factors for some of 
these nuclei. Our results are comparable to those values. For example,
our calculated value for S of $^{212}$Po is 0.019 as compared to theoretical
value 0.025 deduced in \cite{th}.
A value of 0.031 was obtained by Mohr\cite{Mohr} in a double folding model 
calculation using density from experimentally known charge distribution.

The spectroscopic factors for different chains of isotopes generally 
follow a trend though there are exceptions. However, we find that 
expressed in terms of the Casten factor $P=N_pN_n/(N_p+N_n)$, the points,
with some exceptions, show a smooth trend
when the $N_p$ and $N_n$ values are, or in other words, the core is, chosen 
suitably. As earlier pointed out, this method has already been
followed \cite{Casten1,Casten4}.  However, the implications of 
such choices of $N_p$ and $N_n$ values are usually not always obvious.
We will invoke a number different of subshells in various nuclei that are 
designed to produce a
smooth trend and then give microscopic justifications for these choices later in the
paper.

\begin{figure}
\includegraphics{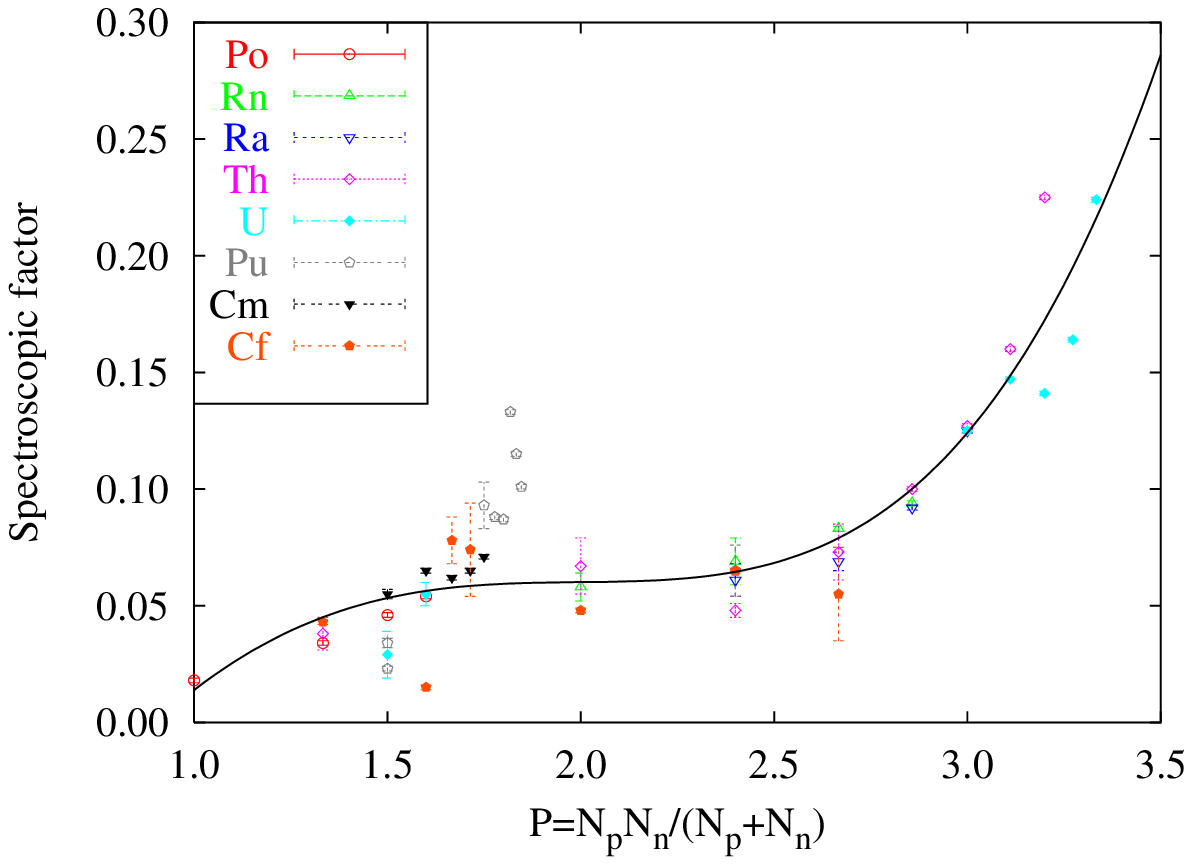}
\caption{Spectroscopic factors $S$ as a function of  $P=N_pN_n/(N_p+N_n)$  for
even-even nuclei between Z=84 and 98.  The $N_p$ and $N_n$ values for
different isotopes are explained in the text. 
\label{figP1}}
\includegraphics{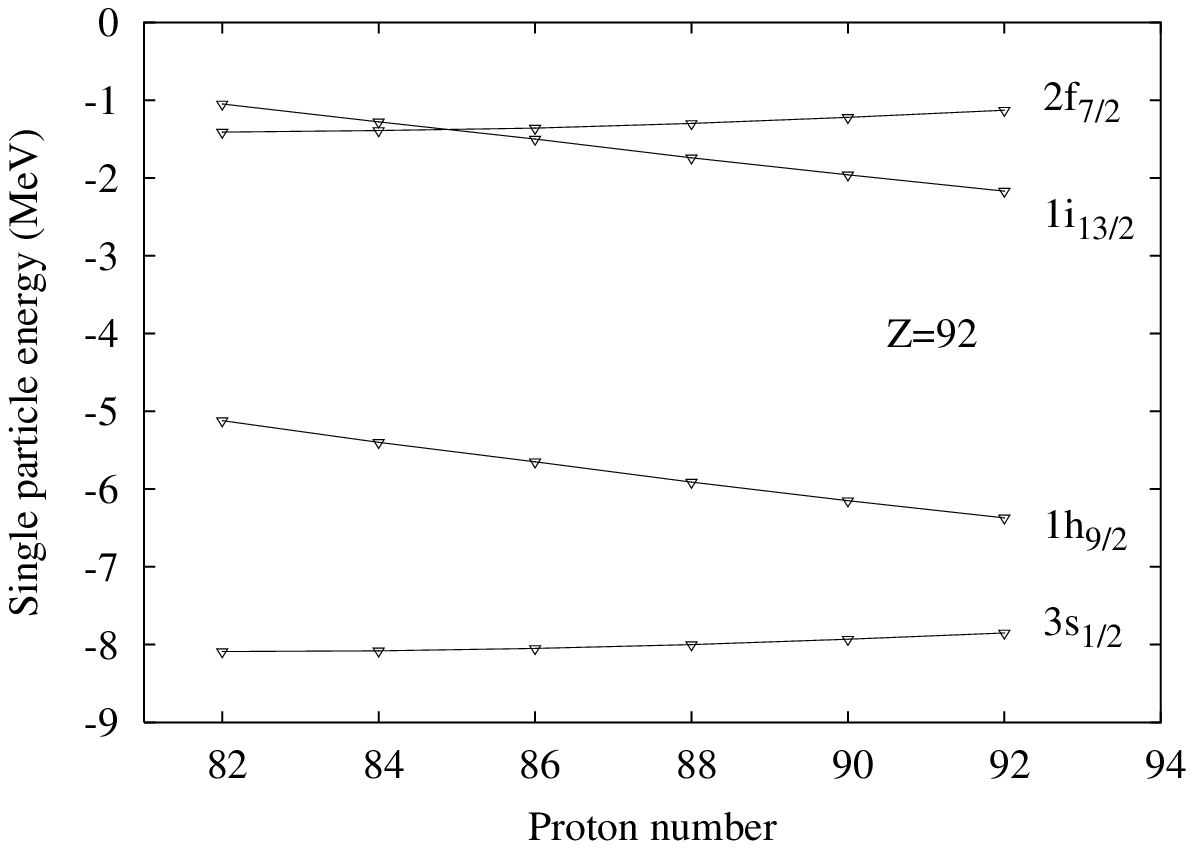}
\caption{ Calculated single particle proton levels near the proton shell closure Z=82
for nuclei with $N-Z=44$.\label{lev}}
\end{figure}

For Po($Z=84$) and Rn(Z=86)
 isotopes, the masses are very close to the doubly magic core $Z=82, N=126$. 
Thus $N_p$ and $N_n$ are chosen simply as $N_p=Z-82$ and $N_n=N-126$, respectively. 
However, the values for Ra($Z=88$) isotopes fall on the smooth curve only if we choose $N_p=4$. This can be justified on the assumption of $Z=92$ as 
a semiclosed shell as explained later.
Similarly, for Th($Z=90$) nuclei, the  effective value needed 
to be chosen is also  $N_p=4$. U($Z=92$) nuclei present a somewhat different 
scenario. For the two lightest U nuclei studied, i.e. $^{224,226}$U, we choose 
$N_p=2$ while for all the heavier isotopes, $N_p=4$ is taken as the effective 
valence proton number. The valence neutron number is taken to be $N_n=N-126$ in all
the above nuclei. For Cm($Z=96$) and Cf($Z=98$) isotopes, we choose
$N_p=2$ and 4, respectively. For Cm isotopes, the valence neutron number
is taken to be $N_n=N-138$ while For Cf isotopes, it is taken as $N_n=152-N$. However 
Pu ($Z=94$) nuclei could not be fitted in the scheme. Interestingly, 
this is analogous to the case of some nuclei with N=90-92 which could not be fitted 
to the $N_pN_n$ scheme for B(E2) values\cite{Casten4} and for the ratio between the excitation energies of the first 4$^+$ and 2$^+$ states\cite{Casten2}. 
The resulting
plot for spectroscopic factors as a function of $P(\ge 1.0)$ is shown in Figure 
\ref{figP1}. The smooth line is drawn only for the 
purpose of guiding the eye. The results for Pu have also been plotted in Figure
\ref{figP1}.
As results for spectroscopic factor
of Pu did not fall on the smooth line in Figure \ref{figP1}, it is difficult
to assign $N_p$ and $N_n$ values in the present procedure. We have assigned 
$N_p$=2 and $N_n=N-126$ for all the Pu nuclei. 

It will be interesting to trace the origin of the effective number of 
valence particles in different isotopes.
As already explained, for Po and Ra nuclei, the number of valence particles
is simply calculated as those outside the $^{208}_{82}$Pb core. Our 
calculation suggests that the next proton single particle level is 
$\pi1h_{9/2}$.
This fact is also experimentally verified by the ground state spin-parity of 
the odd proton nuclei with $Z=83-91$. Particularly, the ground states of the
odd mass nuclei $^{187-215}$Bi($Z=83$), $^{197-217}$At($Z=85$), 
$^{201-219}$Fr($Z=87)$, $^{207-219}$Ac($Z=89$), $^{213-221}$Pa($Z=91$) all have
spin-parity  9/2$^-$, whenever known.  Theoretically, we find 
an interesting situation. The level just below the 82 shell gap, {\em i.e.}
$\pi3s_{1/2}$ and the three levels above it, i.e. $\pi1h_{9/2}$, $\pi2f_{7/2}$
and $\pi1i_{13/2}$ are shown in Figure \ref{lev} for nuclei with $N-Z=44$. It 
can be 
easily seen that the $\pi1h_{9/2}$ state is going down sharply with increasing
$Z$, thus opening up another gap at Z=92. 
Earlier mean field calculations of
Rutz {\em et al.}\cite{Rutz} also suggested a shell gap at $Z=92$ in the vicinity
of nuclei with $N=126$.
Hence, in Ra nuclei protons are
hole-like and $N_p$ is taken as 4. 

In nuclei with higher Z, there is a possibility of another gap opening up near 
$Z=94$. Nuclei with $Z=84-88$ that we have studied have small deformation 
in their ground state. This is evident from the measured B(E2) values. However, 
near $Z=92$, nuclei have large deformation. There is a possibility of new gaps 
opening up in this region for deformed nuclei. 
We should 
mention here that deformed shell gaps are not usually employed in the $N_pN_n$ scheme.
In an earlier calculation
with RMF, Long {\em et al.}\cite{Long} have shown that in nuclei with deformation
$\beta\sim 0.3$ and with $Z$ and $N$ near 96 and 144, respectively,
 gaps of the order of 1.5-2 MeV appear at $Z=94,96$.
There is a similar example in the neutron sector.
Rare-earth nuclei such as Hf, Lu, etc are known to exhibit a deformed shell 
gap  at $N=94$.  This possibly is the reason that Cm and Cf isotopes
have $N_p=2$ and 4, respectively while Th with $Z=90$ has 4 proton holes. 
However, our calculation does not include the effects of deformation and it 
is difficult to be absolutely certain of this explanation. This fact also 
also may explain why $^{224,226}$U have $N_p=2$ but fails for the 
heavier U nuclei where it was necessary to assign $N_p=4$. 

A similar situation also exists in the neutron sector. In nuclei
with neutron number $N>126$, the first level beyond $N=126$ is calculated to be 
the $\nu 1i_{11/2}$ state. In lighter nuclei this state is close to the other 
states in the same shell.
However, as neutron number increases, it comes down in energy rapidly
opening a gap between it and the other states in the same shell. For 
example, in $^{234}$Cm, this gap is more than 3.8 MeV, a value more than double 
of that observed in $^{218}$Po. Thus in Cm, $N_n=138$ may act a 
subshell closure and the number of valence nucleons come out as 
$N=138$. This cannot however explain why for Cf isotopes, it is necessary
to have the number of valence neutrons at $N_n=N-152$. This may possibly
be another effect of deformation. 

\begin{figure}
\includegraphics{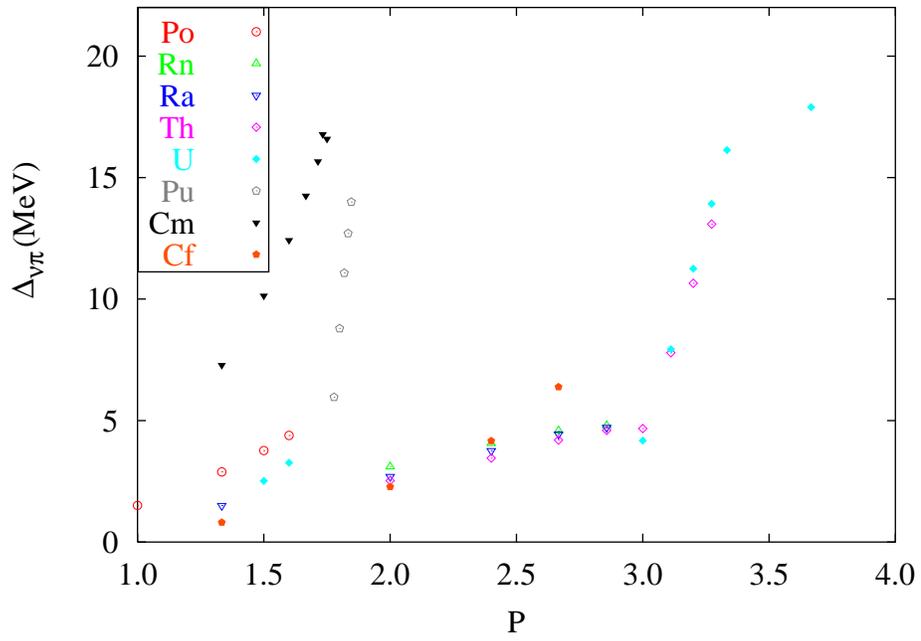}
\caption{The quantity $\Delta_{\nu\pi}$ plotted as a function of $P$  for
even-even nuclei between Z=84 and 98.  See text for details.
\label{be}}
\end{figure}

We expect that the observed smooth behaviour of the spectroscopic factor as a 
function of the parameter $P$ should be reflected in some other properties also
if the parameter $P$ basically is a measure of proton-neutron interaction. 
With this in mind, we have chosen to study the ground state binding energy
of the nuclei in this region. Our method is as follows. 
It is well known that correlations beyond mean field results are due principally
to residual two body interaction. The residual interaction between similar 
nucleons is described by the zero range pairing force in the present
calculation. However, the residual $n-p$ interaction has not been considered in
the present calculation. So for a chain of isotopes, the difference
between the calculated and the experimental binding energies may be a 
measure of the strength of $n-p$ interaction in a particular nucleus. However, 
not all differences
can be ascribed to this effect. We have chosen one nucleus for each Z where,
our scheme suggests that the  number of valence neutrons is zero {\em i.e.}
there is no valence $n-p$ pair. We expect the
effect of $n-p$ interaction to be small in these nuclei and the difference between
the calculated and experimental binding energies in these nuclei to be due to 
all the other effects combined. The difference in the change in the binding 
energy from 
the isotope with $N_n=0$ for a particular Z between theory and experiment is 
taken as a measure of the contribution of $N_pN_n$ interaction and expressed as 
$\Delta_{\nu\pi}$. This difference has been plotted as function of $P$ in Figure
\ref{be} in a scatter diagram. One can see a similar trend as observed in the 
case of spectroscopic factors particularly in the horizontal trend for low $P$
values and the sharply upsloping values beyond $P=3$. The results for Pu are 
somewhat ambiguous for two reasons. 
Firstly, as has already been pointed out, the $N_p$ and $N_n$ values
are somewhat arbitrary.
Secondly, the experimental binding
energy value for Pu with $N_n=0$ {\em i.e.} $N=126$ is not known and we had to 
use an extrapolated value.
  In this case also, Pu 
nuclei do not conform to the general trend. However, the Cm isotopes also do
not follow it. We have checked that the 
B(E2) values  for $2^+\longrightarrow0^+$ also show a similar type of behaviour,
indicating that the quadrupole deformation starts to increase beyond $P=3$, 
consistent with the fact that the proton neutron interaction is the major cause 
of deformation.

To summarize, microscopic superasymmetric fission model has been employed to 
calculate the lifetime of alpha decay in even-even nuclei with $Z=84-98$ and 
$N=128-152$. The interaction 
between the alpha particle and the daughter nucleus has been calculated in 
the double folding model using a density dependent NN interaction. The
densities have been obtained  using the Relativistic Mean Field formalism 
with the Lagrangian density FSU Gold.  
The spectroscopic factors for the decays are shown to
vary smoothly as a function of effective numbers of valence nucleons,
$N_p$ and $N_n$. However, for a useful description, it was often
necessary either to change the usual spherical subshell definitions or
to utlize deformed subshells in counting $N_p$ and $N_n$. The proton level $\pi h_{9/2}$ and the
neutron level $\nu i_{11/2}$ occur just after the closed shells at $Z=82$ and 
$N=126$, respectively and act sometimes as subshells.
Binding energy systematics also follow a similar smooth curve
indicating the role of the residual $n-p$ interaction in spectroscopic factor.
In view of the observed smooth behaviour, in future it will be interesting to
investigate the origin of the observed effective valence nucleon numbers
in more detail, particularly around $Z=94$ and $N=152$.. 

This work was carried out with financial assistance of the
Board of Research in Nuclear Sciences, Department of Atomic Energy (Sanction
No. 2005/37/7/BRNS).

\newpage
\begin{table}
\caption{Logarithm of experimental half life values and spectroscopic factors 
(S) of alpha decay obtained in the present 
calculation.\label{tab1}}
\center
\begin{tabular}{cccccc}\hline
Parent & \multicolumn{1}{r}{log T$_{ex}$(s)} & S & Parent & 
\multicolumn{1}{c}{log T$_{ex}$(s)} & S \\\hline
 $^{212}$Po &  -6.524 & 0.019 & $^{232}$U & ~9.337&0.149\\
 $^{214}$Po &  -3.784 & 0.034 &$^{234}$U &12.889&0.142\\
 $^{216}$Po &  -0.839 & 0.046 &$^{236}$U & 14.868& 0.166\\
 $^{218}$Po &   ~2.269 & 0.055&$^{238}$U & 17.149& 0.226\\
 $^{216}$Rn &  -4.347 & 0.059 &$^{228}$Pu &0.041 & 0.015\\
 $^{218}$Rn &  -1.456 & 0.070 &$^{230}$Pu &2.084 & 0.079\\
 $^{220}$Rn &   ~1.745 & 0.084&$^{232}$Pu &3.997 & 0.075\\
 $^{222}$Rn &   ~5.519 & 0.095&$^{234}$Pu &5.723 & 0.094\\
 $^{220}$Ra &  -1.745 & 0.063 &$^{236}$Pu &7.955 & 0.089\\
 $^{222}$Ra &   ~1.580 & 0.070&$^{238}$Pu &9.442 & 0.088\\
 $^{224}$Ra &   ~5.500 & 0.092&$^{240}$Pu &11.316& 0.135\\
 $^{226}$Ra &  10.703 & 0.125 &$^{242}$Pu &13.073& 0.115\\
 $^{218}$Th & -6.931  & 0.039& $^{244}$Pu &15.402& 0.102\\
 $^{220}$Th & -5.013  & 0.068&$^{240}$Cm &~6.370 & 0.055\\
 $^{222}$Th & -2.552  & 0.049&$^{242}$Cm & ~7.148& 0.066\\
 $^{224}$Th &  ~0.021  & 0.074&$^{244}$Cm & ~8.756& 0.062\\
 $^{226}$Th &  ~3.263  & 0.101& $^{246}$Cm & 11.176 & 0.066\\
 $^{228}$Th &  ~7.780  & 0.128&$^{248}$Cm & 13.078& 0.072\\
 $^{230}$Th & 12.376  & 0.162& $^{242}$Cf & ~2.443 &0.056\\
 $^{232}$Th & 17.646  & 0.225&$^{244}$Cf & ~3.066 & 0.065\\
 $^{224}$U & -3.046&0.030&$^{246}$Cf&~5.109&0.048\\
 $^{226}$U &  -0.456 & 0.055 &$^{248}$Cf& ~7.460&0.043\\
 $^{230}$U & ~6.255&0.127 &$^{250}$Cf&~8.615&0.041\\
\hline
\end{tabular}
\end{table}

\newpage

 
\end{document}